\documentclass[a4paper,fleqn,usenatbib]{mnras}

\usepackage{amsmath}	

\usepackage[T1]{fontenc}
\usepackage{ae,aecompl}

\usepackage{graphicx}	

\usepackage{amssymb}	

\title[(N/O) ratios in the Local Universe]{Nitrogen and oxygen abundances in the Local Universe}
\author[F. Vincenzo et al.]{F. Vincenzo$^{1,2}$\thanks{E-mail:
vincenzo@oats.inaf.it}, F. Belfiore$^{3,4}$, R. Maiolino$^{3,4}$, F. Matteucci$^{1,2,5}$, P. Ventura$^{6}$
\\
$^{1}$Dipartimento di Fisica, Sezione di Astronomia, Universit\`a di Trieste, via G.B. Tiepolo 11, 34100, Trieste, Italy\\
$^{2}$INAF, Osservatorio Astronomico di Trieste, via G.B. Tiepolo 11, 34100, Trieste, Italy\\
$^{3}$Cavendish Laboratory, University of Cambridge, 19 J. J. Thomson Avenue, Cambridge CB3 0HE, UK\\
$^{4}$Kavli Institute for Cosmology, University of Cambridge, Madingley Road, Cambridge CB3 0HA, UK\\
$^{5}$INFN, Sezione di Trieste, Via Valerio 2, 34100, Trieste, Italy\\
$^{6}$INAF, Osservatorio Astronomico di Roma, via Frascati 33, I-0004, Monte Porzio Catone (RM), Italy}

\begin{document}

\date{Accepted 2016 March 1. Received 2016 March 1; in original form 2015 November 19}

\pagerange{\pageref{firstpage}--\pageref{lastpage}} \pubyear{2015}

\maketitle

\label{firstpage}


\begin{abstract}
We present chemical evolution models aimed at reproducing the observed (N/O) vs. (O/H) abundance pattern of star forming galaxies in 
the Local Universe. 
We derive gas-phase abundances from SDSS spectroscopy and a complementary sample of low-metallicity dwarf galaxies, making use of 
a consistent set of abundance calibrations. 
This collection of data clearly confirms the existence of a plateau in the 
(N/O) ratio at very low metallicity, followed by an increase of this ratio up to high values as 
the metallicity increases. This trend can be interpreted as due to two main sources of nitrogen in galaxies: 
i) massive stars, which produce small amounts of pure primary nitrogen and are responsible for the (N/O) ratio in the 
low metallicity plateau; ii) low- and intermediate-mass stars, which produce both secondary and primary nitrogen and 
enrich the interstellar medium with a time delay relative to massive stars, and cause the increase of the (N/O) ratio.  
We find that the length  of the low-metallicity plateau is almost solely determined by the star formation efficiency, which regulates 
the rate of oxygen production by massive stars. We show that, to reproduce the high observed (N/O) ratios at high (O/H),  as well as the right slope of the 
(N/O) vs. (O/H) curve, a differential galactic wind -- where oxygen is assumed to be lost more easily than nitrogen -- 
is necessary. 
No existing set of stellar yields can reproduce the observed trend without assuming differential galactic winds.
Finally, considering the current best set of stellar yields, a bottom-heavy initial mass function is favoured to reproduce the data. 
\end{abstract}




\section{Introduction} 
\label{sec:intr} 

The self-regulation of star formation in galaxies by gas accretion and galactic outflows is a fundamental ingredient in the modern framework of galaxy evolution. 
Both the analysis of large galaxy surveys and the direct observation of high-velocity clouds 
(e.g. \citealt{fraternali2002,oosterloo2007,heald2011,gentile2013}), as well as hydrodynamical simulations  
\citep{prochaska2009,faucher2011,vandevoort2011,dekel2013,fraternali2015} 
point towards the need for continuous accretion of  pristine gas onto galaxies (see also \citealt{fraternali2008,putman2012}). Moreover, historically, 
 the so-called 
 `G-dwarf problem' \citep{vandenbergh1962,schmidt1963,tinsley1980} was solved by relaxing the hypothesis of a closed-box evolution of the solar neighborhood and 
 allowing the accretion of pristine gas onto the disk, 
 which acts in diluting the abundances; in this way, one can reconcile the predicted frequency 
 of metal-poor Galactic disk stars (which are too many in the framework of the simple closed-box model) with the observed one. 
Large scale galactic winds and outflows are also necessary to reproduce the overall properties of the observed galaxy population and 
to match the observed chemical enrichment of the intergalactic medium (IGM; see, for example, \citealt{finlator2008,erb2008,fabian2012,hopkins2012}). 
Several observations of galactic outflows in the literature have demonstrated the 
ubiquity of the outflow phenomenon, both locally and at high redshift (see, for example, 
\citealt{pettini2001,bolatto2013,geach2014,erb2015,cicone2014,cicone2015}), 
however understanding their impact on galaxy properties over cosmic time remains a 
daunting task.

\par Since metals are a direct product of star formation in galaxies, chemical abundances are a powerful probe of the feedback processes driving the evolution of galaxies. 
Oxygen occupies a key role in this type of studies, since its gas phase abundance can be inferred from strong nebular lines easily observed in the optical 
wavelength 
range in the low redshift Universe. Since oxygen is the most common metal by mass, its abundance is also an excellent proxy for the total metallicity of the gas. Moreover, since 
oxygen is mostly produced by massive stars, dying as core-collapse supernovae (SNe), its enrichment is relatively simple to model and does not require, to first approximation, taking into 
account the effect of stellar lifetimes (under the so-called instantaneous recycling approximation, IRA).

\par The study of the relation between oxygen abundance and other fundamental galaxy parameters like stellar mass \citep{lequeux1979,tremonti2004}, star formation rate 
(SFR, \citealt{mannucci2010,andrews2013}), gas content \citep{hughes2013,bothwell2013} and environment \citep{pasquali2010,peng2014} has paved the way to the development 
of a new generation of chemical evolution models \citep{dave2012,lilly2013,peng2014,lu2015,belfiore2016}, 
which succeed, to various extents, at reproducing the general trends observed in the data with cosmological inflow rates and various simple outflow prescriptions.


\begin{figure*}
\includegraphics[width=0.8\textwidth]{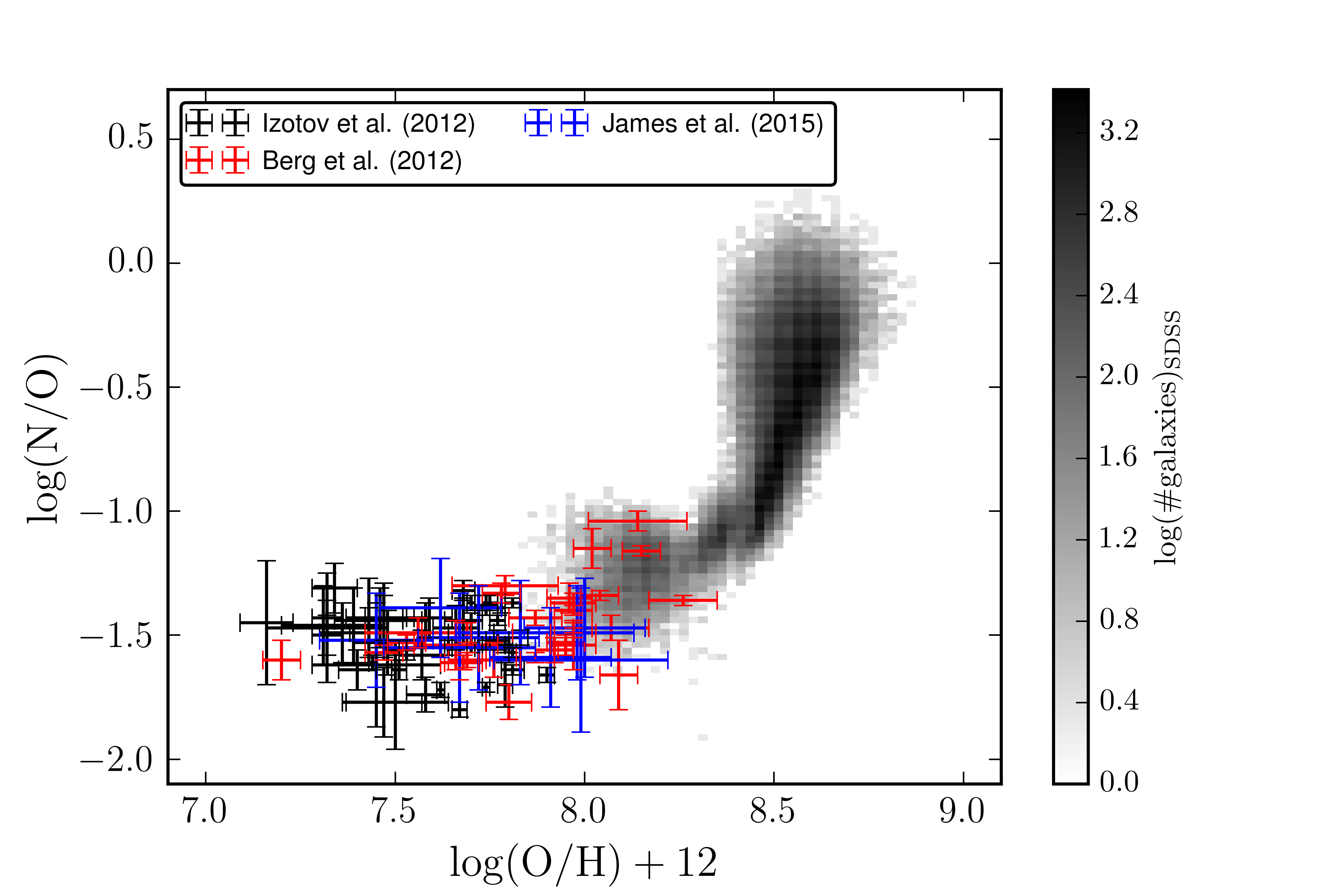}
\caption{ In this figure, the SDSS data sample for the (N/O) vs. (O/H) abundance pattern (density plot in greyscale) 
is compared with a data sample which includes the (N/O) and (O/H) abundances as observed in blue, diffuse star forming dwarf galaxies 
by \citet[red data]{berg2012}, \citet[black data]{izotov2012} and \citet[blue data]{james2015}. 
}
     \label{fig:data_all}
\end{figure*} 


\par Since different chemical elements are released in the interstellar medium (ISM) on different time-scales, the study of abundance ratios of some key elements can provide 
tighter constraints on the star formation and gas flow history of a system.  For example, large galaxy surveys have demonstrated that the [O/Fe] ratios 
in elliptical galaxies are consistent with the paradigm requiring these objects to form stars vigorously over a short timescale at high redshift \citep{matteucci1994,bernardi2003,pipino2004,thomas2005,thomas2010}.  Moreover, detailed analysis of chemical abundance ratios of  n-capture, $\alpha-$ and iron-peak elements 
in individual stars in the Milky Way (MW) 
have been instrumental in showing that the various components of our Galaxy (halo, bulge, thick and thin disk) have had different chemical evolution histories with respect to each 
other (see, for example, \citealt{pagel2009,matteucci2001book,matteucci2012}). 

\par In the context of star forming galaxies, strong nebular lines in the optical range allow reliable measurement of the (N/O) abundance ratio when both the 
[OII]$\lambda3726$,$3729$ and [NII]$\lambda6548$,$6584$ doublets are measured. The (N/O) abundance ratio has been studied by several authors 
\citep{vila-costas1992,thuan1995,henry2000,chiappini2003,chiappini2005,perez-montero2009,perez-montero2013,belfiore2015} since it is a promising tool to 
shed light on the relative 
role of pristine gas inflows and outflows, which appear degenerate when only the abundance of one chemical element is traced 
(see \citealt{koppen2005}, but also the discussion in the Appendix of \citealt{belfiore2016}).

\par Unlike oxygen, nitrogen is a chemical element mostly produced by low- and intermediate-mass stars (LIMS), with the nucleosynthetic yields depending 
on metallicity in a complex fashion. 
In particular, a stellar generation can release into the galaxy ISM both primary and secondary N. 
The secondary N component increases with metallicity, being a product of the CNO cycle and formed at expenses of the C and O 
already present in the star. Concerning LIMS, the primary N component is produced during the third dredge-up, occurring 
along the asymptotic giant branch (AGB) phase, 
if nuclear burning at the base of the convective envelope is efficient \citep{renzini1981}. 
The latter is particularly important for very metal-poor LIMS, which would not be capable otherwise of synthesizing significant amounts of secondary N. 
On the other hand, the computation of the N stellar yields for massive stars still suffers of large uncertainty, especially at very low metallicity, 
and none of the current existing stellar evolutionary codes 
is able to provide the right amount of primary N which is needed to reproduce the observed (N/O) plateau at very low metallicities. 

\par All the aforementioned complications prevent one from using IRA  
and a constant yield for the study 
of nitrogen abundances in galaxies, 
and have prevented the community so far from taking full advantage of the large nitrogen abundance datasets now available both through large spectroscopic surveys of local 
(like SDSS, \citealt{york2000}, or GAMA, \citealt{driver2011}) and high-redshift (e.g. zCOSMOS, \citealt{lilly2009}) galaxies. 

In this work we critically revise the different assumptions affecting the interpretation of the (N/O) versus (O/H) abundance patterns, making use of a large dataset of star forming galaxies 
from the SDSS, complemented by data from metal-poor dwarf galaxies to explore the low-metallicity regime. While considerable uncertainties still persist in some of the basic model 
parameters (nitrogen yields, stellar initial mass function, IMF) we aim here at setting new constraints on pristine gas inflows and the 
models of outflows. 

The paper is structured as follows. In Section \ref{sec:data}, we present the dataset used in this work for the (N/O) vs. (O/H) abundance diagram. 
In Section \ref{sec:nucl}, we summarize the current knowledge about the nucleosynthetic origin of nitrogen in stars. 
In Section \ref{sec:descr}, we describe the basic equations and the assumptions of the numerical model of chemical evolution adopted in this work. 
Our results are presented in Section \ref{sec:sdss_model} and \ref{sec:low_Z_tail}; in particular, in Section \ref{sec:sdss_model}, we focus on the results of our models for 
the (N/O) vs. (O/H) abundance pattern of the ensemble of the SDSS galaxies; 
in Section \ref{sec:low_Z_tail}, we present the results of our models for the low metallicity plateau, which a complementary sample of metal-poor, diffuse and 
star-forming dwarf galaxies exhibit in the (N/O) vs. (O/H) diagram. Finally, in Section \ref{conclusions}, we end with our conclusions. 

\section{Overview of the data} 
\label{sec:data}

In order to study the characteristic gas-phase oxygen and nitrogen abundances in local galaxies, we make use of the data from SDSS Data Release 7 \citep[DR7]{abazajian2009} 
and the emission line fluxes, stellar masses and SFR estimates presented in the MPA-JHU catalogue\footnote{The MPA-JHU catalogue is available 
online at \url{https://www.sdss3.org/dr8/spectro/galspec.php} } \citep{brinchmann2004,tremonti2004,kauffmann2003} released as part of DR 8 
\citep{aihara2011}.

\par We select galaxies to have $0.023<z<0.3$ and S/N$>$3 on the following emission lines: [OII]$\lambda\lambda3726$,$28$; [OIII]$\lambda5007$; H$\beta$; H$\alpha$; 
[NII]$\lambda6584$, and [SII]$\lambda\lambda6717$,$31$. We use the standard Baldwin-Philipps-Terlevich (BPT) diagram \citep{baldwin1981,veilleux1987,kauffmann2003} 
to exclude sources where the gas ionisation is not dominated by star formation, since available metallicity calibrations are only tailored to star forming regions. In this work we use the 
[OIII]/H$\beta$ versus [SII]/H$\alpha$ diagnostic diagram and the demarcation curve of \citet{kewley2001}. We do not make use of the popular [OIII]/H$\beta$ versus [NII]/H$\alpha$ 
to avoid a bias against nitrogen enriched H\textsc{ii} regions \citep{sanchez2015,belfiore2015}.

\par Emission line fluxes are then corrected for dust extinction using the Balmer decrement and the \citet{calzetti2001} reddening curve with $\rm R_V = 4.05$. The theoretical value for 
the Balmer line ratio is taken from \citet{osterbrock2006}, assuming case B recombination ($ \mathrm{H \alpha  / H \beta=2.87 }$). 
We note that the use of extinction curves of \citet{cardelli1989} or \citet{charlot2000} yield very similar results for the wavelength range considered in this work.

\par Inferring gas phase oxygen abundance from strong nebular line ratios is a difficult problem, since the line ratios depend not only on ionic abundances, but also on other parameters, such as ionisation parameter, density and hardness of ionisation field. It is well known in the literature that different oxygen abundance calibrations based on strong nebular lines can lead to systematic discrepancies of up to $0.6$ dex \citep{kewley2008,lopezsanchez2012,blanc2015}.  In particular, strong line calibrations based on an extrapolation to high metallicity of abundances  
measured with the $\tau_{\mathrm{e}}$ method (which makes use of the faint oxygen auroral line [OIII]$\lambda4363$ 
to directly infer the electron temperature of the nebula) generally lead to lower 
abundances than calibrations based on photoionisation models. 

Moreover, several metallicity calibrators make use of the nitrogen line fluxes, thus implicitly assuming that the relationship between the (N/O) ratio 
and metallicity varies monotonically with oxygen abundance.

\par In this work, we infer the oxygen and nitrogen abundances using the self-consistent framework presented in 
\citet{pilyugin2010}, which calibrates various strong line ratios through the electron temperature method. An alternative calibration taking both oxygen and nitrogen abundance into account has been recently presented in \citet{perez-montero2014}.

\par In order to sample the low metallicity regime, which is poorly populated in SDSS, we make use of the data from \citet{izotov2012,berg2012,james2015} for a collection of blue, diffuse and star forming dwarf galaxies. We note that the abundances reported by \citet{izotov2012,berg2012,james2015} correspond to the chemical abundances as measured using the direct method and hence should fall onto the same scale of chemical abundances we inferred from the SDSS data with the adopted calibration.

\par In Fig. \ref{fig:data_all} we show the trend of the observed (N/O) ratios as a function of the (O/H) abundances. The density plot in greyscale 
represents the abundance pattern as 
observed in the ensemble of the SDSS galaxies, whereas the data with error bars represent the compilation of star forming dwarf galaxies from \cite{berg2012,izotov2012,james2015}. The latter data extend towards lower (O/H) abundances than the SDSS data and clearly exhibits the well-known 
low metallicity plateau. The SDSS data show an abrupt change of the slope at oxygen abundances higher than $\rm 12+log(O/H) \sim 8.4\,\mathrm{dex}$ .


\subsection{Estimating the dust depletion} \label{subsec:dust}

Oxygen abundance calibrations based on nebular lines only trace the oxygen abundance of the gaseous phase of the ISM. 
However, chemical elements in real galaxies are partially depleted on to dust grains. Since chemical evolution models only 
predict the total metallicity, depletion onto dust grains must be taken into account when comparing models with our data.
From an observational point of view, the dust content can differ among galaxies of different metallicity and SFR \citep{dacunha2010,fisher2014,hjorth2014}. 
In the framework of chemical evolution studies, 
the most important physical process affecting the dust cycle in galaxies is the star formation history, which regulates the main feedback processes responsible for the 
dust production and destruction and hence the run of the galaxy dust-to-gas ratio with metallicity 
\citep{wang1991,dwek1998,lisenfeld1998,edmunds2001,calura2008,dwek2011,feldmann2015}. 


\begin{figure*}
\label{fig:lims_yield}
\includegraphics[width=14cm]{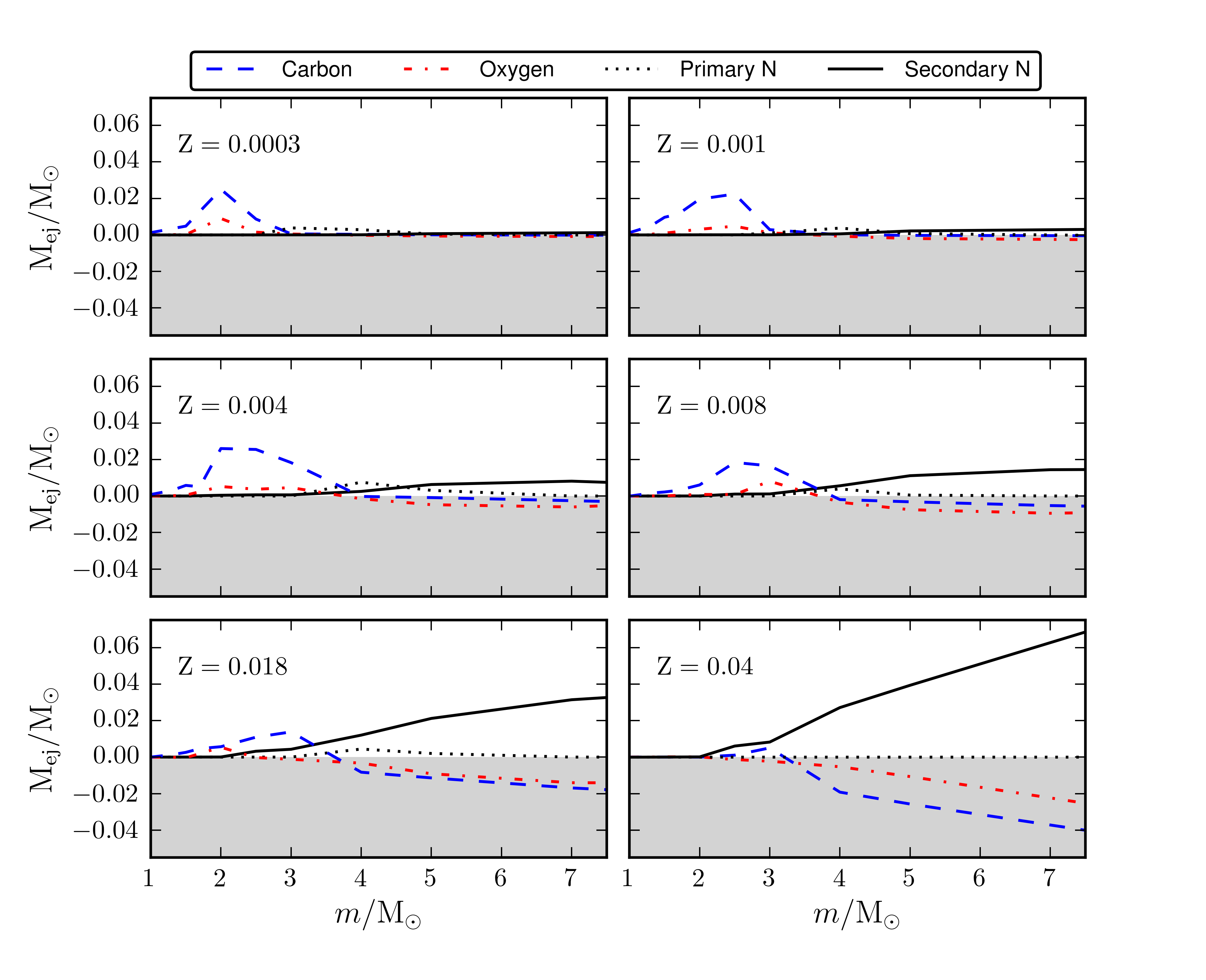}
\caption{In this figure, we show the \citet{ventura2013} stellar yields of LIMS for carbon (dashed blue line), oxygen (dashed-dotted red line), primary nitrogen (dotted black line) 
and secondary nitrogen (solid black line), as computed at $Z=0.0003,\,0.001,\,0.004,\,0.008,\,0.018$, and $0.04$. The various quantities do not include the amount of ejected mass which was initially 
present in the star and has not been nuclearly processed; so the finding of negative values for the stellar yields means that the final total ejected amount of the generic element $X$ is 
smaller than the one which was initially present in the star at its birth and has been later ejected into the ISM without any nuclear processing.}
\end{figure*}


\par In the literature a mean oxygen depletion\footnote{The depletion factor $D(X)$ is the logarithmic decrement 
between the observed abundance of a chemical element and its predicted total abundance, 
$A(X)$, namely $ D(X) \equiv \log(N_{X}/N_{H})_\text{obs} - A(X)$. 
} of $D(\text{O}) \approx -0.1\,\text{dex}$ is often assumed (see also \citealt{whittet2003,jenkins2009,whittet2010}). 
Although oxygen depletion is likely to have a dependance on metallicity 
(differential depletion), for simplicity we do not consider this effect in this work.
Although \cite{jenkins2009} suggests an average nitrogen depletion of $\sim-0.1\,\mathrm{dex}$, 
the large uncertainty in this estimate means that nitrogen is also consistent with zero depletion. Indeed most of the studies indicate that 
nitrogen is not a refractory element and 
does not deplete onto grains even in the densest molecular clouds (e.g. \citealt{meyer1997,caselli2002}). 
In light of this, in our work, we do not consider any nitrogen depletion onto dust. 
While the depletion corrections applied in this work are rather arbitrary, none of our conclusions depend on the exact values of the depletion factors.


\section{The nucleosynthetic origin of nitrogen} 
\label{sec:nucl}
 
According to its mass and initial chemical composition, each star pollutes the ISM with different amounts of a given chemical element. Since stellar lifetimes 
primarily depend upon the stellar mass, each chemical element is expected to enrich the ISM of galaxies on different typical timescales.  

\subsection{Primary and secondary nucleosynthetic products}
\label{sec:primary}

\par A fundamental aspect to take into account in chemical evolution models is the nature of the physical processes which give rise to the production of the various chemical elements in stars.  
In particular:
\begin{enumerate}
\item If a sequence of nuclear reactions involves as initial seed only the H and He present in the gas mixture of the star at its birth, then the nucleosynthetic 
products of that sequence do not depend upon the initial stellar metallicity. The chemical products of these reactions are then named \textit{primary} elements.
\item If the presence of metals in the initial gas mixture of the star is necessary for some nuclear reactions to occur, then the 
nucleosynthetic products of those nuclear reactions depend on the metallicity. These chemical products are named \textit{secondary} elements. 
\end{enumerate} 
In modern chemical evolution studies, the importance of distinguishing between the secondary and primary nature has been simplified for most of the chemical 
elements, since their yields are computed as a function of the initial metallicity. 
Nevertheless, for chemical elements like nitrogen, which have both a primary and secondary origin (see, for example, \citealt{edmunds1978,alloin1979,renzini1981,matteucci1986,gavilan2006,molla2006}), 
it can be conceptually useful to separate the two components, even in the presence of yields computed for different metallicities.

\subsection{The production of nitrogen in the CNO cycle}

\par Nitrogen is mainly produced during the CNO cycle, whose main branch consists in a series of $p$-captures and $\beta^{+}$ decays starting from an atom of $^{12}$C 
and converting four protons into a nucleus of He (with two $e^{+}$ and two $\nu_{e}$ as byproducts). 
Since the inner reaction $^{14}\text{N}(p,\gamma)^{15}\text{O}$ proceeds with the slowest rate among the other in the sequence, when the equilibrium condition is reached 
(namely, when the rate of production of each CNO nucleus equals its rate of destruction), the ultimate effect of the CNO cycle is to convert most of the CNO isotopes 
into $^{14}$N. The origin of the initial $^{12}$C seed in the CNO cycle is a discriminating factor. In fact, if the atom of $^{12}$C was initially present in the gas 
mixture from which the star originated, then the synthesized nitrogen behaves as a secondary element. On the other hand, if some physical mechanism is able to carry 
the C and O nuclei produced in the He-burning zones out to the H-burning zones, then the synthesized nitrogen behaves like a primary element.

\subsection{Nitrogen yields in low- and intermediate-mass stars}

\par Low- and intermediate-mass (LIM) stars during the AGB phase eject into the ISM significant amounts of He, C and N. 
The ejected masses reflect important abundance variations. The main physical mechanism for the transport of the C and O nuclei to the outer zones 
of the star is given by the so-called third dredge-up, which is the transfer of the nucleosynthetic products of the triple-$\alpha$ process by the surface 
convection which proceeds after each thermal pulse. Primary nitrogen can then be produced when the hot-bottom burning (CNO burning at the base of the 
convective envelope) occurs in combination with the third dredge-up (see, for example, \citealt{ventura2013}).

\par In Fig. \ref{fig:lims_yield}, we show how the stellar yields of \citet{ventura2013} for C, O, and primary and secondary N vary as a function of the initial stellar mass, 
for different metalliticities. The stellar yield of a given chemical element $X$ is defined as the ejected amount of mass of the newly formed $X$. By looking at the figure, 
the stellar yields of secondary N increase with metallicity, by means of the consumption of the C and O nuclei originally present in the star. In fact, during the CNO cycle, 
the global abundance of the CNO nuclei remains constant, while the relative abundances of each CNO element can significantly vary. By looking at Fig. \ref{fig:lims_yield}, 
the stellar producers of primary N have masses in the range between $\sim3\,\text{M}_{\sun}$ and $6\,\text{M}_{\sun}$. Furthermore, the production of primary N does not show 
any correlation with the consumption of the original C and O in the star, as expected. 
 

 \begin{table*}
 \caption[]{ \footnotesize{In this Table, the various columns report the following quantities: 
 i) Z, metallicity; ii) $\overline{M}_{\mathrm{O,R10}}$, the IMF-averaged stellar yield of oxygen 
in the mass range $M=11$-$100\,\mathrm{M}_{\sun}$, by assuming the stellar yields of \citet[R10]{romano2010} and 
the \citet{kroupa1993} IMF; iii) $\overline{M}_{\mathrm{N,R10}}$, the IMF-averaged stellar yield of nitrogen, defined as in the 
previous column; iv) predicted average $\log(\text{N/O})$ ratios, when considering only the contributions of massive stars; 
v) average stellar yield of primary N which should be provided by massive stars 
to reproduce the observed low-metallicity plateau with $\log(\text{N/O})\approx-1.6\,\text{dex}$. This empirically-derived yield is used as the reference stellar yield for primary nitrogen 
at low metallicity in this work.
}}

 \begin{tabular}{c | c c c | c }
 \hline
  Z & \small{$\overline{M}_{\mathrm{O,R10}}$ [$\text{M}_{\sun}$] } & \small{$\overline{ M}_{\mathrm{N,R10}}$ [$\text{M}_{\sun}$] } & \small{$\log(\text{N/O})_\text{R10}$} 
  & \small{$\overline{M}_{\text{N,prim}}$  [$\text{M}_{\sun}$]} \\
 \hline
\rule{0pt}{4ex}
  $1.0\times10^{-10}$ &  $3.025$ &  $0.014$ &  $-2.27$ &  $0.066$   \\
 \rule{0pt}{2.5ex}   
  $1.0\times10^{-8}$ &  $3.105$ &  $0.082$ &  $-1.52$ &  $0.068$   \\
 \rule{0pt}{2.5ex}   
   $1.0\times10^{-5}$ &  $1.995$ &  $0.0005$ &  $-3.53$ &  $0.043$   \\
 \rule{0pt}{2.5ex}   
   $1.0\times10^{-3}$ &  $1.999$ &  $0.005$ &  $-2.55$ &  $0.044$   \\
 \rule{0pt}{2.5ex}   
   $4.0\times10^{-3}$ &  $2.013$ &  $0.018$ &  $-1.99$ &  $0.044$   \\
 \hline
\end{tabular}

\label{table_yields}
\end{table*} 

 
\subsection{Nitrogen yields in massive stars} 

\par In order to reproduce the observed plateau in the (N/O) ratio for the MW halo stars, \citet{matteucci1986} proposed that massive 
stars should produce only primary N at all metallicities, at variance with standard nucleosynthesis models predicting only secondary N 
from massive stars. 
This plateau was also later observed in damped Ly$\alpha$ (DLA) systems \citep{pettini2002,pettini2008} and blue, low 
metallicity star forming dwarf galaxies \citep[see Fig. \ref{fig:data_all}]{thuan1995,izotov2012,berg2012,james2015}, thus confirming that 
there should be primary N production in massive stars. 

\par \citet{maeder2000} and \citet{meynet2002b} found that primary N can be produced in rapidly rotating massive stars, 
but only at very low metallicity. 
Their proposed main physical mechanism is given by the so-called rotational mixing, particularly efficient at very low metallicity, 
where stars are expected to rotate faster and to be much more compact than their metal-rich counterparts. 
Rotational mixing allows the nucleosynthetic products of the 
triple-$\alpha$ reaction (mainly C and O nuclei) to efficiently diffuse towards the outer CNO burning zone, where a pure primary nitrogen 
can then be synthesized  \citep{chiappini2008, maeder2009}. 

\par This unfortunately cannot solve the problem of the observed (N/O) plateau,  
which extends all over the metallicity range of MW halo stars. In fact, Geneva stellar evolutionary models predict 
fast rotating massive stars to produce primary N only in a narrow range at very low metallicity; 
as the metallicity $Z>10^{-8}$, these stars resume producing only secondary N.  
Therefore, by assuming the theoretical stellar yields of the Geneva group, 
chemical evolution models predicted a `dip' in the (N/O) ratio 
at low (O/H) abundances, which is never observed (see, for example, \citealt{chiappini2005}, and the further discussion in 
Section \ref{sec:empirical_yield} and \ref{sec:low_Z_tail}). 
To solve this discrepancy and be able to explain the observed trends in the data, 
all previous chemical evolution models \citep{matteucci1986,chiappini2005} 
had to assume artificially a pure primary N yield from massive stars.  

\subsection{Empirically fixing the nitrogen yield from massive stars at low metallicity}
\label{sec:empirical_yield}

In order to re-assess the problem of primary nitrogen production in massive stars, we suggest a revised primary N stellar yield at low metallicity. 

\par In Table \ref{table_yields} we report the IMF-averaged stellar yields of N and O from massive stars, $\overline{M}_{\mathrm{N,R10}}$ and $\overline{M}_{\mathrm{O,R10}}$, respectively, 
as computed by using the \citet{romano2010} compilation of stellar yields and assuming the \citet{kroupa1993} IMF. 
We also report how the values of $\log(\text{N/O})$ are predicted to vary for different metallicities, when considering only the contributions of massive stars to the nitrogen and oxygen 
chemical enrichment of the ISM. 
We observe that the Geneva stellar models predict a dip in the (N/O) ratios for $\mathrm{Z}>10^{-8}$. This dip is never observed and it stems 
from the lack of primary N production by massive stars for $\mathrm{Z}>10^{-8}$. 

 \par In order to reproduce the low-metallicity plateau, we therefore calculate the average amount of primary N that massive stars should provide to 
 reproduce the observed (N/O) plateau 
 at very low metallicities. We assume that the oxygen yields in this regime are reliable and we require the predicted (N/O) ratios to match the observed  
 $\log(\text{N/O})\approx-1.6\,\mathrm{dex}$ at low metallicity. 
In particular, we empirically derive the needed average stellar yield of primary nitrogen by massive stars as 
\begin{equation} \label{eq:Nprim}
\overline{M}_{\text{N,prim}}(\text{Z})= \overline{M}_{\text{O,R10}}(\text{Z})\times10^{-1.6}\times\frac{A_\text{N}}{A_\text{O}},
\end{equation} where $A_\text{N}$ and $A_\text{O}$ represent the atomic weight of N and O, respectively. 
The calculated values of $\overline{M}_{\text{N,prim}}(\text{Z})$ are reported in the last column of Table \ref{table_yields}. Interestingly, at extremely low Z, 
the latter are of the same order of magnitude as the value of $0.065\,\text{M}_{\sun}$ originally adopted in chemical evolution models to reproduce the low-metallicity plateau (see also \citealt{chiappini2005}, and references therein).  

\par For the rest of this work, in our reference chemical evolution models, we assume a pure primary N production by massive stars, 
with the stellar yields at the various metallicities being the quantity $\overline{M}_{\text{N,prim}}(\text{Z})$, as given in Eq. (\ref{eq:Nprim}) and reported in the last column of Table \ref{table_yields}. 
We remark on the fact that, because of the way we have defined it, this quantity depends on the assumed IMF and stellar yields of oxygen.

\section{The chemical evolution framework} \label{sec:descr}

In this work, we study the nitrogen and oxygen evolution in the ISM of galaxies by adopting a chemical evolution model 
in which the galaxy is assumed to be composed of a single zone within which the various chemical elements are assumed to mix instantaneously and uniformly. 
The basic ingredients of the model are described in detail in \citet{matteucci2012}. In summary, the model is capable of following the temporal evolution of 
the abundances of various chemical elements within the ISM of galaxies, by taking into account the main physical processes driving chemical evolution, 
such as star formation, inflows and outflows of gas.

\subsection{Star formation and chemical evolution}

\par By defining $M_{\text{g},i}(t)$ as the gas mass in the galaxy which is in the form of the $i$-th chemical element at time $t$, its temporal evolution follows: 

\begin{multline} 
\label{eq:chem_mod} 
 \frac{\text{d}{M}_{\text{g},i}(t)}{\text{d}t} = \underbrace{-X_{i}(t)~\text{SFR}(t)}_{\text{SF}} + \underbrace{R_{i}(t)}_{\text{yields}} - 
\underbrace{\Psi_{i}(t)}_{\text{outflow}} + \underbrace{\Phi_{i}(t)}_{\text{infall}}, 
\end{multline} where $X_{i}(t)=M_{\text{g,}i}(t)/M_{\text{g}}(t)$ is the abundance by mass of the $i$-th chemical element, defined such that $\sum_{i}{X_{i}(t)}=1$, and $M_{\text{g}}(t)$ is the total gas mass in the galaxy at time $t$. In our models, we assume the stellar lifetimes of \citet{padovani1993}. 

\par We assume a star formation law of the form $ \text{SFR}(t)=\nu M_{\text{g}}(t)$, with $\nu$ being the star formation efficiency (SFE), a free parameter of our models.
\par The term $R_{i}(t)$ in equation \ref{eq:chem_mod} represents the rate at which stars return the $i$-th chemical element back to the ISM at their death. 
This term subsumes all our prescriptions about the stellar yields as well as the assumptions concerning SN progenitors. 
In particular, for Type Ia SNe, we assume the `single degenerate scenario' with the same prescriptions as in \citet{matteucci2001, matteucci2001book,matteucci2012}, however the details of the treatment of Type Ia SNe are largely irrelevant to this work, since they only have a very minor effect in the enrichment of oxygen and nitrogen.  


\subsection{The infall rate}

\par In our model, the galaxy is assumed to assemble by means of accretion of gas from an external reservoir 
into the potential well of an underlying dark matter (DM) halo. 
 
\par The gas infall rate, $\Phi_{i}(t)$, follows an exponential form with time given by
\begin{equation} 
 \Phi_{i}(t)=\frac{X_{\text{inf},i}\,M_{\mathrm{inf}}\,e^{-t/\tau_{\text{inf}}}}{\tau_{\text{inf}}\left( 1-e^{-t_{\text{G}}/\tau_{\text{inf}}} \right)}, \;\text{with} 
\sum_{i}{\int_{0}^{t_{\text{G}}}{\Phi_{i}(t)dt}}=M_{\text{inf}}, 
\end{equation} where $t_{\text{G}}$ is the age of the galaxy and $X_{\text{inf},i}$ is the abundance by mass of the $i$-th chemical element in the infalling gas ($M_{\text{inf}}$), 
whose chemical composition is assumed to be primordial.


\subsection{The outflow model}

In equation \ref{eq:chem_mod} the outflow rate is modelled by $\Psi_{i}(t)=\omega_{i}~\text{SFR}(t)$, where $\omega_{i}$ is the so-called \textit{mass loading factor}. 
Observations generally suggest that star-forming galaxies experience time-averaged outflow loading factors of order unity \citep{lilly2013,peng2015,belfiore2015,lu2015} 
at stellar masses around $\log(M_{\star} / M_{\sun}) \sim 10$, close to the knee of the luminosity function. It is, however, likely that less massive galaxies experience much higher loading factors.

\par In this work, we assume the galactic wind to be \textit{differential}; namely, the outflow carries only the main nucleosynthetic products of core-collapse SNe (mainly $\alpha$-elements), 
for which $\omega_{i}$ is a constant value. For chemical elements such as nitrogen and carbon, which have a very minor contribution from SNe, we assume a null mass loading factor, 
i.e.  $\omega_{i}=0$. The assumption of a differential outflow is justified by the fact that massive stars (the progenitors of core-collapse SNe 
and the most important oxygen producers in the Universe) are observed to be highly clustered and so, as they explode, they create a region of the ISM in which the filling factor closely approaches to unity (see \citealt{marconi1994,recchi2001}). 


\begin{figure*}
\includegraphics[width=0.8\textwidth]{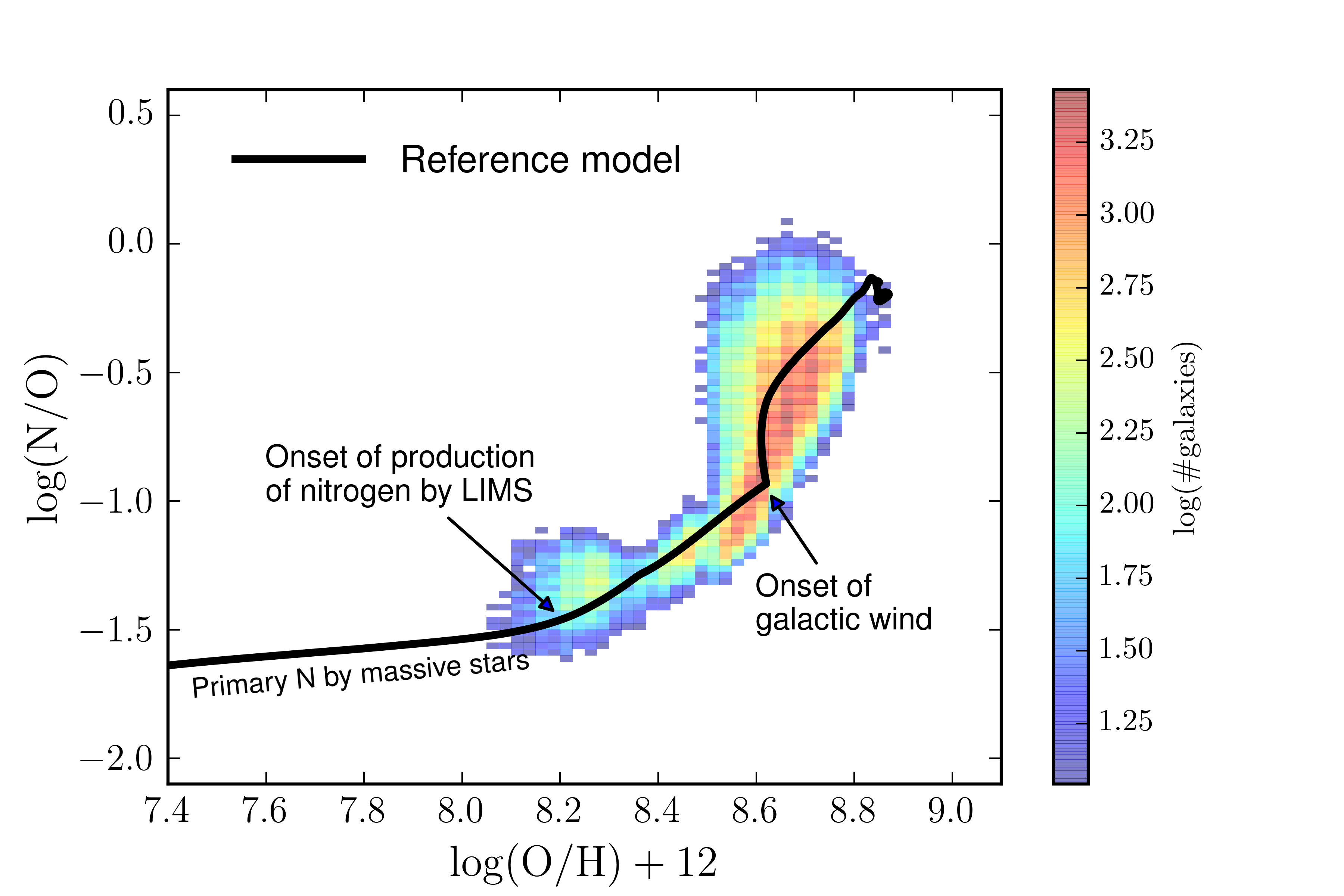}
\caption{The reference chemical evolution model used in this work to reproduce the abundance pattern in the (N/O) vs (O/H) plane (black solid line). The distribution of SDSS galaxies is shown 
as a 2D histogram, with the bin size in both the (N/O) and (O/H) dimensions being $0.025\,\mathrm{dex}$, and the colour-coding represents the number of galaxies in each bin. 
The changes in slope of the observed (N/O) vs (O/H) relation are linked to different physical properties of the model. The plateau at low metallicity is due to the pure primary N production 
by massive stars. The slope of the (N/O) vs (O/H) relation increases when LIMS start dying, producing both primary and secondary nitrogen. A further steepening of the relation is 
obtained after the onset of a differential galactic wind, which here is assumed to expel oxygen preferentially. }
\label{fig:ref_model}
\end{figure*} 


Interestingly, the first works suggesting a differential metal-enhanced galactic wind in the context of the study of chemical evolution of galaxies were those of \citet{pilyugin1993} and \citet{marconi1994}, which addressed also the issue of explaining the (N/O) vs. (O/H) abundance pattern observed in dwarf irregular galaxies (see \citealt{recchi2008} for a detailed study and references). 
We are aware that our assumption of a differential galactic wind is highly uncertain and it does not rely on firm 
theoretical and observational findings; further detailed investigations are needed, by looking -- for example -- at the chemical abundances in the halo of the 
MW or in quasar absorption lines, which were also related in the past to galactic winds.   

\par Following the formalism of \citet{bradamante1998}, the time for the onset of the galactic outflow is calculated by requiring the thermal energy of the gas (supplied by SNe 
and stellar winds to the galaxy ISM) to be larger than the binding energy of the gas to the galaxy potential well.


\subsection{Summary of the stellar yields for O and N}

In this work we assume for oxygen and nitrogen the following set of stellar yields.
\begin{enumerate}

\item For massive stars, we assume the metallicity-dependent compilation of stellar yields of \citet{romano2010}, in which the nitrogen and oxygen yields have been computed 
by the Geneva group, by including the combined effect of rotation and mass loss (see, for more details, \citealt{meynet2002,hirschi2005,hirschi2007,ekstrom2008}). 
For $ \text{Z} < 10^{-3}$ we make use of the empirically-motivated nitrogen yield of massive stars derived in Sec. \ref{sec:empirical_yield}. 

\item For low- and intermediate-mass (LIM) stars, we assume the stellar yields at the various metallicities computed by means of the ATON numerical code of stellar evolution 
(see, for a detailed description, \citealt{mazzitelli1989,ventura1998, ventura2009, ventura2013}). 
We have chosen the \citet{ventura2013} stellar yields because they provide separately the primary and secondary components of the nitrogen stellar yield for a large range of 
metallicities ($3.0\times10^{-4}\leq \text{Z} \leq 0.04$; see the discussion in Sec \ref{sec:primary}). 
Other works which also separate the two components are those of \citet{gavilan2005,gavilan2006}, however they span a too narrow metallicity 
range for the purpose of our work. 

\end{enumerate}



\begin{figure*}

\includegraphics[width=\textwidth]{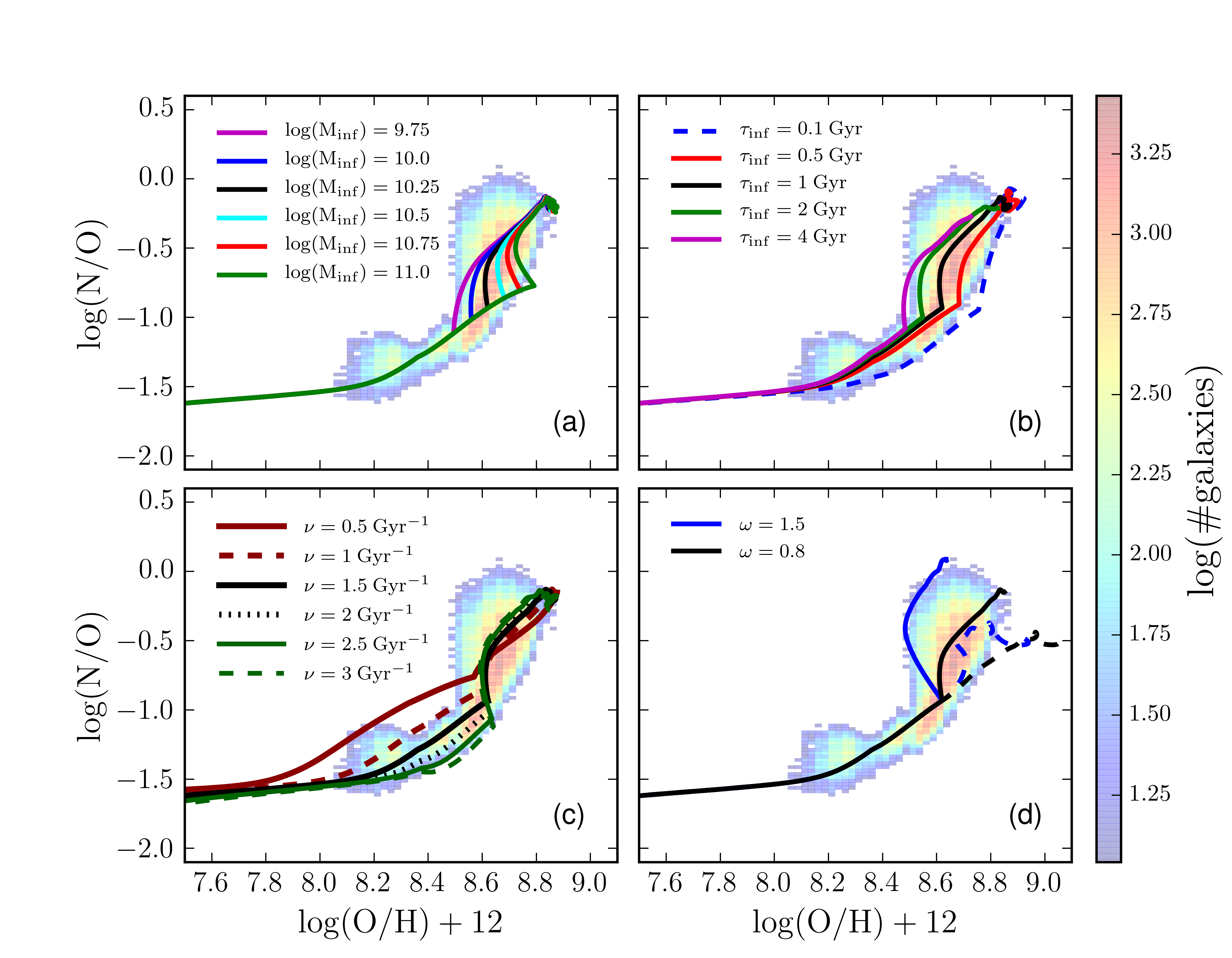}
\label{fig:param}

\caption{Chemical evolution model track in the (N/O) vs (O/H) plane. Abundances for SDSS galaxies are shown as a 2D histogram, with the colour-coding corresponding 
to the number of galaxies in each bin, with the bin size in both the (N/O) and (O/H) dimensions being 0.025 dex. 
The reference model is always plotted in black solid line. 
In different panels we vary different free parameters of the chemical evolution model. In panel a) we change the infall mass $\rm M_{inf}$, 
in panel b) the infall timescale $\rm \tau_\mathrm{inf}$, in panel c) the SFE $\nu$ and in panel d) the outflow loading factor $\omega$. 
In panel d) the dashed lines correspond to a non-differential outflow (where both N and O are expelled with the same efficiency) while the solid lines refer to the reference 
assumption of a differential outflow where N is not expelled (with $\omega_{N}=0$). }

\end{figure*}


\section{Modelling the SDSS data} 
\label{sec:sdss_model}

In this section, we compare the predictions of our chemical evolution models with the set of data discussed in Sec. \ref{sec:data} (see Fig. \ref{fig:data_all}) focusing on reproducing the high-metallicity regime of the (N/O) versus (O/H) diagram. We will address the question of reproducing the low metallicity tail of the (N/O) distribution in the next section.

\subsection{The reference model}
\label{sec:reference}

\par Given the relatively large number of free parameters in our detailed chemical evolution model, we take a qualitative, staged approach. We first choose a set of reference parameters. We then investigate the effect of changing each of them, while keeping the other ones fixed. 
Our reference model assumes:
\begin{enumerate}
\item fixed SFE $\nu=1.5\,\text{Gyr}^{-1}$,
\item fixed mass loading factor $\omega=0.8$ for oxygen, 
\item differential galactic outflow, with $\omega=0.0$ for nitrogen,
\item infall mass $\log(\text{M}_\text{inf} / \text{M}_{\sun})=10.25$, 
\item infall time-scale $\tau_\text{inf}=1\,\text{Gyr}$,
\item \citet{kroupa1993} IMF,
\end{enumerate}
and the star formation history is assumed to be extended over all the galaxy lifetime (i.e. we do not introduce a quenching phase).
In Fig. \ref{fig:ref_model} we plot this reference model in the (N/O) versus (O/H) plane, together with the distribution of SDSS galaxies. The observed (N/O) vs. (O/H) abundance pattern is shown as a density plot, with the bin size in both the (N/O) and (O/H) dimensions being $0.025\,\mathrm{dex}$; we show only the bins which contain more than ten galaxies and the colour-coding represents the number of galaxies within each bin.

\par It is remarkable to note that our simple reference model in Fig. \ref{fig:ref_model} can well reproduce the main features of the data. 
The predicted low-metallicity plateau is the effect of our assumption of a pure primary N production by massive stars. 
The increase of the (N/O) ratios from $12+\log(\text{O/H}) \sim 8.0\,\mathrm{dex}$ is due to the delayed chemical feedback of LIMS, 
which pollute the ISM with primary \textit{and} secondary nitrogen; we will refer to this change in slope as the `first break point'. 
By definition, the production of secondary nitrogen by LIMS increases as the metallicity increases. 
Although the production of primary N is smaller than the one of secondary N (see Fig. \ref{fig:lims_yield}), 
the main primary N producers turns out to be, on average, less massive and hence more long-living than the bulk of the secondary N producers. 
In this way, the pollution of the ISM with primary N by LIMS mimics and amplifies the secondary N component. 
At $12+\log(\text{O/H}) \sim 8.6\,\mathrm{dex}$, we see in Fig. \ref{fig:ref_model} a new change in slope (`second break point'), 
which is caused by the onset of the galactic wind. 
Since we assume a differential outflow, the loss of oxygen per unit time is more efficient than the loss of nitrogen (which is set to zero). 
In this way, the accumulation of oxygen within the galaxy ISM slows down and the net effect of the galactic wind is to steepen the (N/O) ratios 
as the chemical evolution proceeds. We remark that the increase of the (N/O) ratios after the second break point is also crucially bolstered by 
the larger amounts of secondary N which LIMS are able to synthesize at higher Z.

\subsection{Exploring the parameter space}

In this section we discuss the effect of varying the free parameters in the reference model. In particular, 
we show that the main parameters influencing the shape of 
the (N/O) vs (O/H) relation are the star formation efficiency, the outflow loading factor and the assumptions 
regarding the differential outflow loading for N and O. 
Other parameters, such as the infall mass and the infall timescale do not have a significant effect on the abundance trends studied.

\subsubsection{The infall mass}

In Fig. \ref{fig:param}a), we explore the effect of varying the infall mass, in the range $\log(\text{M}_\text{inf}/\text{M}_{\sun})=9.75$-$11.0$. 
The common evolution of all the galaxies 
before the onset of the galactic wind stems from the fact that all these models assume a fixed SFE ($\nu=1.5\,\text{Gyr}^{-1}$). 
The models developing the galactic wind first are the ones with the smallest infall mass, which depart from the common track at the lowest (O/H) abundances; 
in fact, such models are characterized by a lower galaxy potential well, and hence they develop the outflow at earlier times. 
Overall the infall mass in our models does not play a key role in defining the shape of the (N/O) vs (O/H) relation. 

\subsubsection{The infall timescale}

The infall time-scale regulates the rate of accretion of the gas into the system. In particular, by fixing the values of the other parameters, 
models with longer infall time-scales predict the galactic wind to develop at earlier times, since the binding energy of the gas to the whole galaxy is lower, 
at any time of the galaxy evolution. 
This can be appreciated by looking at Fig. \ref{fig:param}b), where models assuming different infall time-scales are compared. 
Before the onset of the galactic wind, all the models evolve on the same track in the (N/O) vs. (O/H) diagram because 
the SFE is kept fixed.

\subsubsection{The star formation efficiency}

The SFE ($\nu$) is a key parameter in driving the star formation history of the system and has a complex effect on the balance of the different stellar populations that contribute 
to the nitrogen and oxygen abundance in galaxies. 
Increasing the SFE speeds up the production of oxygen per unit time by massive stars in the earliest stages of the galaxy chemical evolution. 
Since we assume that nitrogen is synthesized by massive stars as a pure primary element, 
the increasing of the SFE does not affect much the (N/O) ratio in the low metallicity regime.

\par In Fig. \ref{fig:param}c), we explore the effect of varying the SFE in the range $\nu=0.5-3\,\text{Gyr}^{-1}$. 
The first large effect in the (N/O) vs (O/H) diagram can be seen at the first break point; 
in particular, the higher the SFE, the larger is the metal content within the galaxy as first LIMS die, causing the (N/O) ratios to increase. In conclusion, an increase of the SFE 
determines a wider range in metallicity of the initial (N/O) plateau due to the chemical enrichment of massive stars. 
This is a sort of application of the so-called `time-delay model'\citep{tinsley1979,greggio1983,matteucci_greggio1986} to the (N/O) vs. (O/H) diagram. 
We remind the reader that by \textit{time-delay model} we mean the classical way of interpreting the trend of the observed [$\alpha$/Fe] vs. [Fe/H] abundance patterns in galaxies, where 
key roles are played by the assumed SFE and  
the delayed chemical enrichment by Type Ia SNe; in particular, 
the higher the SFE in galaxies, the larger is the [Fe/H] abundance of the ISM as first Type Ia SNe explode and hence 
the [$\alpha$/Fe] ratios steeply decrease. 


\begin{figure}

\includegraphics[width=9.3cm]{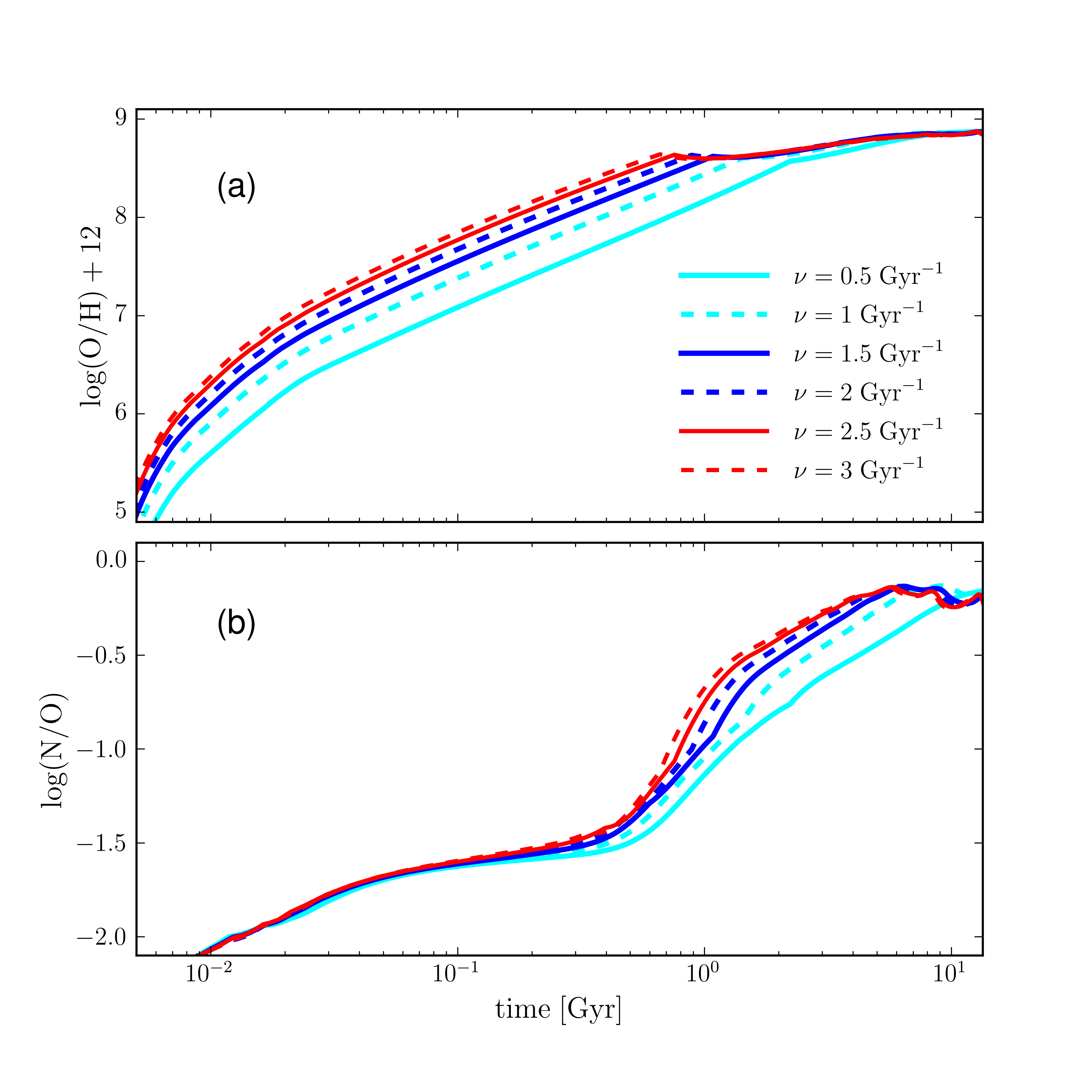}

\caption{ In this figure, in panel a) and b), we show how the (O/H) abundances and the (N/O) ratios, respectively, are predicted to vary as functions of the galaxy time. The various curves 
correspond to models with different star formation efficiency, with the reference model being the blue solid line. }

\label{fig:time_evol}

\end{figure}


\par In Fig. \ref{fig:time_evol}a), we show the relations between the age of the galaxy and the metallicity of the ISM, as predicted by our models with varying SFEs. 
Galaxies with very low SFEs struggle to reach high (O/H) abundances, spending most of their evolutionary time at low metallicities. Conversely, models 
with higher SFEs predict galaxies to reach the (O/H) abundances of the SDSS data at earlier times. 
Once the galactic wind develops, there is an interplay between the rate of restitution of oxygen into the galaxy ISM by dying stars and the rate of removal of 
oxygen by galactic wind and star formation; 
this causes the (O/H) abundances to increase more gradually with time than before the onset of the galactic wind.

\par In Fig. \ref{fig:time_evol}b), we show how the predicted (N/O) ratios vary as a function of the time, when varying the SFE. 
In the earliest stages of the galaxy evolution, all the models tend towards $\log(\text{N/O})\sim-1.6\,\mathrm{dex}$, 
which corresponds to the nitrogen-to-oxygen ratio 
of the low metallicity plateau. 
As first LIMS die, the (N/O) ratio is predicted to suddenly increase, because of the large amounts of both primary and secondary N which LIMS are capable of 
synthesize. In summary, the higher the SFE, the earlier and the higher are the metallicities (as discussed above) when LIMS start dying, 
causing the (N/O) ratio to increase. 
The effect of the onset of the galactic wind is less visible in this figure, and it corresponds to the gradual change in the slope of the (N/O) vs. time relation occurring at later times. 


\begin{figure}
\includegraphics[width=9.3cm]{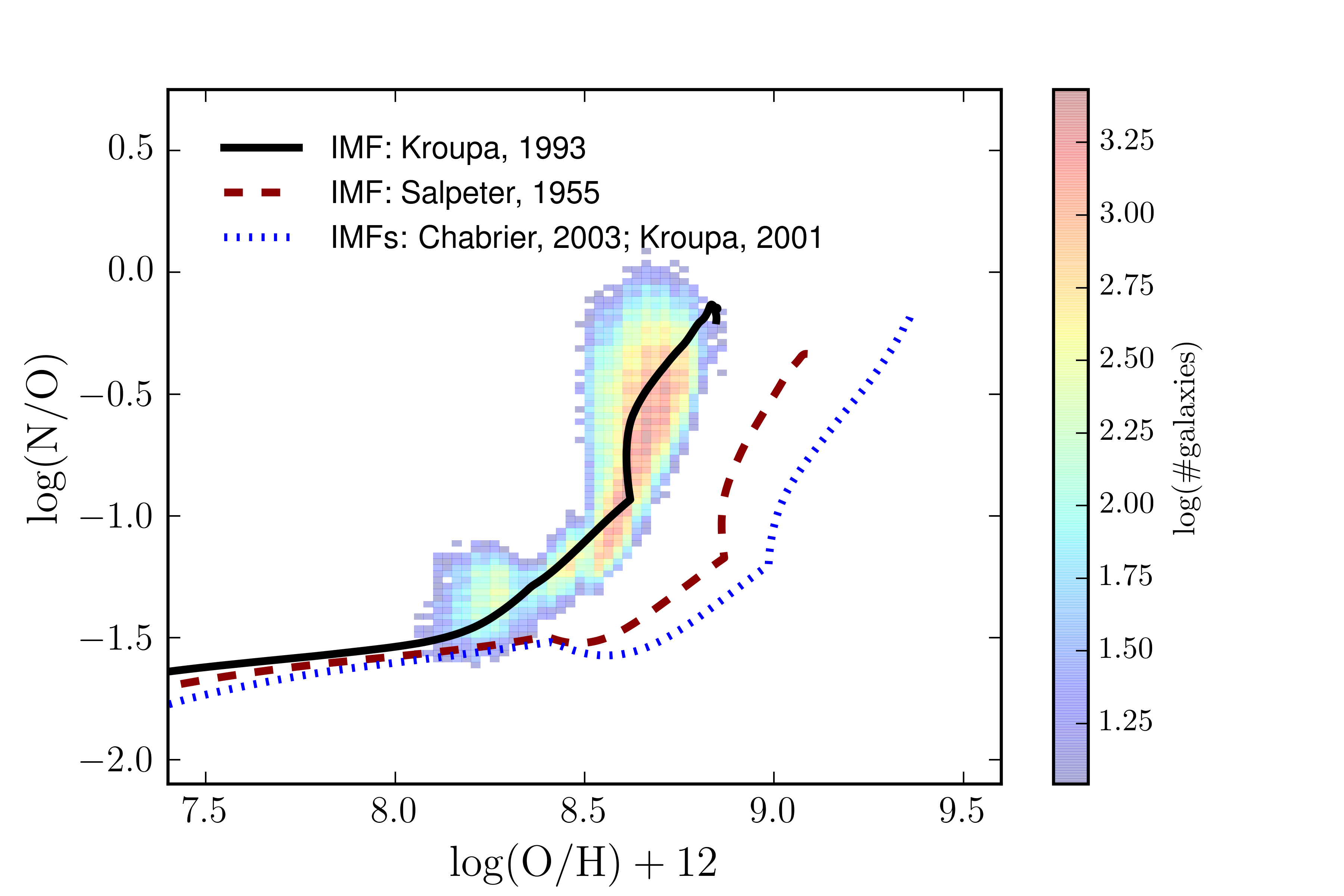}
\caption{ In this figure, we explore the effect of changing the IMF on the predicted (N/O) vs. (O/H) diagram. The black solid curve corresponds to our reference model with 
the \citet{kroupa1993} IMF, the dashed curve in dark red to the \citet{salpeter1955} IMF, and the dotted curve in blue to the \citet{chabrier2003} 
and \citet{kroupa2001} IMFs, which provide very similar final results.
}
     \label{fig:imf}
\end{figure} 


\subsubsection{The outflow loading factor}

Our reference model assumes a differential outflow, which carries only the nucleosynthetic products of core-collapse SNe (mainly $\alpha$-elements) out of the galaxy potential well. 
Hence the mass loading factor, $\omega$, quoted in the reference model only refers to oxygen, since nitrogen is assumed not to be expelled. 
Obviously, in reality, some nitrogen would be lost to the IGM,  but here we consider its mass loading factor to be much lower than the one of oxygen, 
since core-collapse SNe are minor contributors of N in galaxies and the galactic wind (mainly triggered by SN explosions) develops when LIMS have already heavily polluted the ISM with nitrogen, overcoming the N enrichment from massive stars. 


\begin{figure*}
\includegraphics[width=0.8\textwidth]{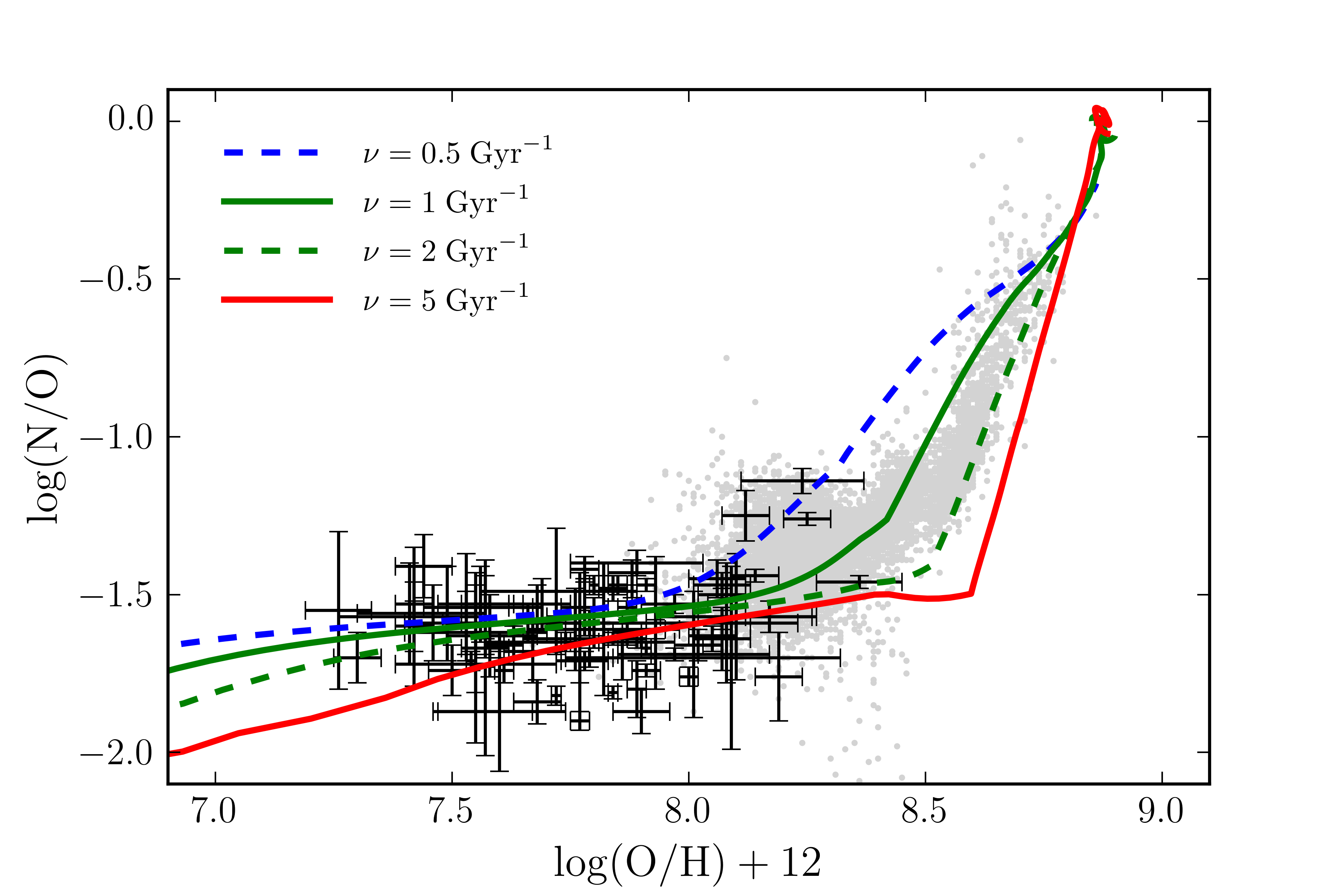}
\caption{ In this figure, we show chemical evolution models for dwarf galaxies. We assume $\text{M}_{\text{inf}}=10^{9}\,\text{M}_{\sun}$, varying SFEs, an infall timescale 
 $\tau_{\text{inf}}=0.1\,\text{Gyr}$ and the IMF of \citet{kroupa1993}. The grey points correspond to the SDSS galaxies with $M_{\star}\le 10^{9}\,{M_{\sun}}$, 
 in the same range of stellar mass as the metal-poor, star forming dwarf galaxies from \citet{berg2012,izotov2012,james2015}, which are represented by 
 the data with error bars. 
 }
 \label{fig:dwarf}
\end{figure*} 


\par In Fig. \ref{fig:param}d), we show the effect of almost doubling the mass loading factor on the (N/O) vs. (O/H) abundance pattern. 
By definition this parameter only has an effect after the galactic wind has started; this 
occurs at $12 + \log(\text{O/H}) \sim 8.6\,\mathrm{dex}$ in our reference model. 
The requirement of a differential outflow to reproduce the observed trend of the (N/O) vs. (O/H) abundance pattern can be appreciated 
by comparing the solid and dashed curves in Fig. \ref{fig:param}d), corresponding to models with differential and non-differential outflow, respectively. 
Concerning the model with a differential outflow, $\omega$ has the clear effect of changing the slope of the (N/O) vs (O/H) relation at high metallicity. 
Importantly, even in the case of the reference model, this slope is much steeper than unity, which is the naive prediction for the secondary nitrogen enrichment. 
By looking at Eq. \ref{eq:chem_mod}, the slope is crucially determined by the balance between the loss of oxygen via galactic winds (and star formation) 
and the restitution of oxygen by massive stars. If the latter exceeds the former, the slope is positive, otherwise it is negative. 
Concerning the model with a non-differential outflow, the transition between the SF-dominated regime and the outflow-dominated regime is smooth, since both oxygen and nitrogen are lost from the 
galaxy potential well with the same efficiency $\omega$.

\subsubsection{The IMF}

In Fig. \ref{fig:imf}, we explore the effect of changing the IMF on the (N/O) vs. (O/H) abundance diagram. 
Our reference IMF, which is \citet[the best IMF for the MW disk; see \citealt{romano2010}]{kroupa1993}, hosts a large number of LIMS and a much smaller number of massive stars 
than the \citet{salpeter1955} IMF, and it provides the best 
agreement with the observed dataset among the classical IMFs considered in this work. IMFs like \citet{chabrier2003} and \citet{kroupa2001}, 
which are very similar among each other, host a larger number of 
massive stars and hence an enhanced oxygen-production is predicted at the early stages of galaxy evolution; so the first break point occurs at higher metallicities 
when the \citet{chabrier2003} or the \citet{kroupa2001} IMF are assumed. 
The main effect of the IMF is to shift the chemical evolution tracks along the (O/H) axis. In fact, the main role of the IMF is to assign different weights to stars in different mass ranges.


\section{Modelling the low metallicity tail} \label{sec:low_Z_tail}

In this subsection, we present chemical evolution models for star forming, metal-poor dwarf galaxies, in order to reproduce the observed low metallicity tail of the (N/O) ratios. The data of \citet{berg2012,izotov2012,james2015} clearly exhibit a plateau in the (N/O) ratios, which extends towards low (O/H) abundances. None of the models developed in the past has been capable of reproducing this trend (see \citealt{chiappini2005}). 

\par In Fig. \ref{fig:dwarf}, we show the predictions of models with $M_{\mathrm{inf}}=10^{9}\,\mathrm{M}_{\sun}$, $\tau_\mathrm{inf}=0.1\,\mathrm{Gyr}$ 
and SFEs in the range $\nu=0.05-5\,\text{Gyr}^{-1}$, with the purpose of reproducing the trend of the (N/O) vs. (O/H) abundances 
which are observed in metal-poor star forming dwarf galaxies (data with error bars) and in the SDSS galaxies with stellar mass $M_{\star}\le10^{9}\,\text{M}_{\sun}$. 
The other parameters are the reference ones, as described at the beginning of  Section \ref{sec:sdss_model}. In these models, we assume that massive stars produce 
pure primary N; our stellar yields for N are summarized in Table \ref{table_yields}, 
and they have been computed as a function of the metallicity, starting from the observational constraint that metal-poor dwarf galaxies share a common 
nitrogen-to-oxygen ratio which is $\log(\text{N/O})\approx-1.6\,\text{dex}$.

\par The trend of the various models in Fig. \ref{fig:dwarf} can be explained by means of the same mechanism which has been mentioned 
throughout all the text: very low SFEs cause a slow production of oxygen by massive stars; 
this fact allows the ISM to be quite metal-poor when LIMS begin to highly pollute the ISM with nitrogen. Therefore, lowering the SFEs causes the plateau of the (N/O) ratios 
to be less extended in metallicity and hence the first break to occur at lower Z. 
This explanation is a sort of application of the time-delay model to the (N/O) vs. (O/H) diagram. 
By comparing models and data, we can obtain a good qualitative agreement, clearly suggesting that a pure primary N production by massive stars 
is needed to explain and reproduce the low-metallicity plateau. 

\par By assuming the N stellar yields of massive stars collected by \citet{romano2010}, which include the results of 
the stellar evolutionary code of the Geneva group taking 
into account the effects of mass loss and rotation, 
all our chemical evolution models predict a dip in the (N/O) ratios when going towards low (O/H) abundances 
(see also Table \ref{table_yields}), 
at variance with observations. 

This can be appreciated by looking at Fig. \ref{fig:dwarf_originalN}, where the predictions of a model with pure primary N production 
by massive stars is compared with a similar model assuming the Geneva stellar yields for massive stars, as given in \citet[their model 15]{romano2010}. 
 Both models assume $\nu=1.5\,\text{Gyr}^{-1}$, $M_{\mathrm{inf}}=10^{9}\,\mathrm{M}_{\sun}$ and $\tau_\mathrm{inf}=0.1\,\mathrm{Gyr}$. Fig. \ref{fig:dwarf_originalN} clearly points out the still open problem of standard stellar nucleosynthesis calculations as well as of stellar evolutionary models 
 to predict the right amount of pure primary N which massive stars should provide to reproduce the observed (N/O) plateau at very low metallicity.


\section{Conclusions} \label{conclusions}

In this article, we have presented a set of chemical evolution models with the purpose of reproducing the (N/O) vs. (O/H) abundance pattern, as 
observed in a sample of SDSS galaxies \citep{abazajian2009} and metal-poor, star forming dwarf galaxies \citep{izotov2012,berg2012,james2015}. 
Our collection of data spans a wide metallicity range ($7.1\,\mathrm{dex} \la \log(\mathrm{O/H})+12 \la 8.9 \,\mathrm{dex}$), enabling us to recover the trend 
of the observed (N/O) vs. (O/H) relation with a precision never reached before.
At very low metallicity, the data clearly demonstrate the existence of a plateau in the (N/O) ratio, 
followed by an increase of this ratio which steepens as the metallicity increases. We summarize our main conclusions in what follows.

\begin{enumerate}

\item The low metallicity plateau in the 
nitrogen-to-oxygen ratio represents the imprint of pure primary N from massive stars, as originally suggested by \citet{matteucci1986}. 
Such plateau is also observed 
all over the metallicity range of MW halo stars and in low metallicity DLAs. 
From a theoretical point of view, standard nucleosynthesis calculations have shown that the rotational mixing 
in very-metal poor massive stars can allow a pure primary N production 
(see \citealt{maeder2000}, and subsequent papers from the Geneva group); 
nevertheless, as the metallicity becomes $Z>10^{-8}$, massive stars resume producing 
only secondary N. This still represents an open problem in
stellar nucleosynthesis calculations, since all the chemical evolution models with the current stellar yields of massive stars, including standard 
mass loss and rotation, predict a ``dip'' in the (N/O) ratios for $\mathrm{Z}>10^{-8}$, 
at variance with observations (see, for example, \citealt{chiappini2005,romano2010}).

\item In this work, we have computed the primary N stellar yields 
of massive stars which are needed, as functions of the metallicity, to reproduce the observational constraint suggesting that 
$\log(\text{N/O})\approx-1.6\,\text{dex}$ in metal-poor, star forming 
dwarf galaxies. Our results are given in the last column of Table \ref{table_yields}. 
In this way, we have been able to reproduce the observed flat trend of the data. 

\item A fundamental aspect to take into account 
for explaining the trend of the observed (N/O) vs. (O/H) abundance pattern is the time-delay with which LIMS start enriching the ISM with both primary 
and secondary N. In fact, before the first LIMS die, massive stars are the only nitrogen and oxygen producers in galaxies. 
When LIMS start dying, the N abundance within the galaxy ISM steeply increases. Since the stellar yields of the secondary N component by LIMS increase with metallicity, then the (N/O) ratios continuously grow as a function of the (O/H) abundance. 


\begin{figure}
\includegraphics[width=9.3cm]{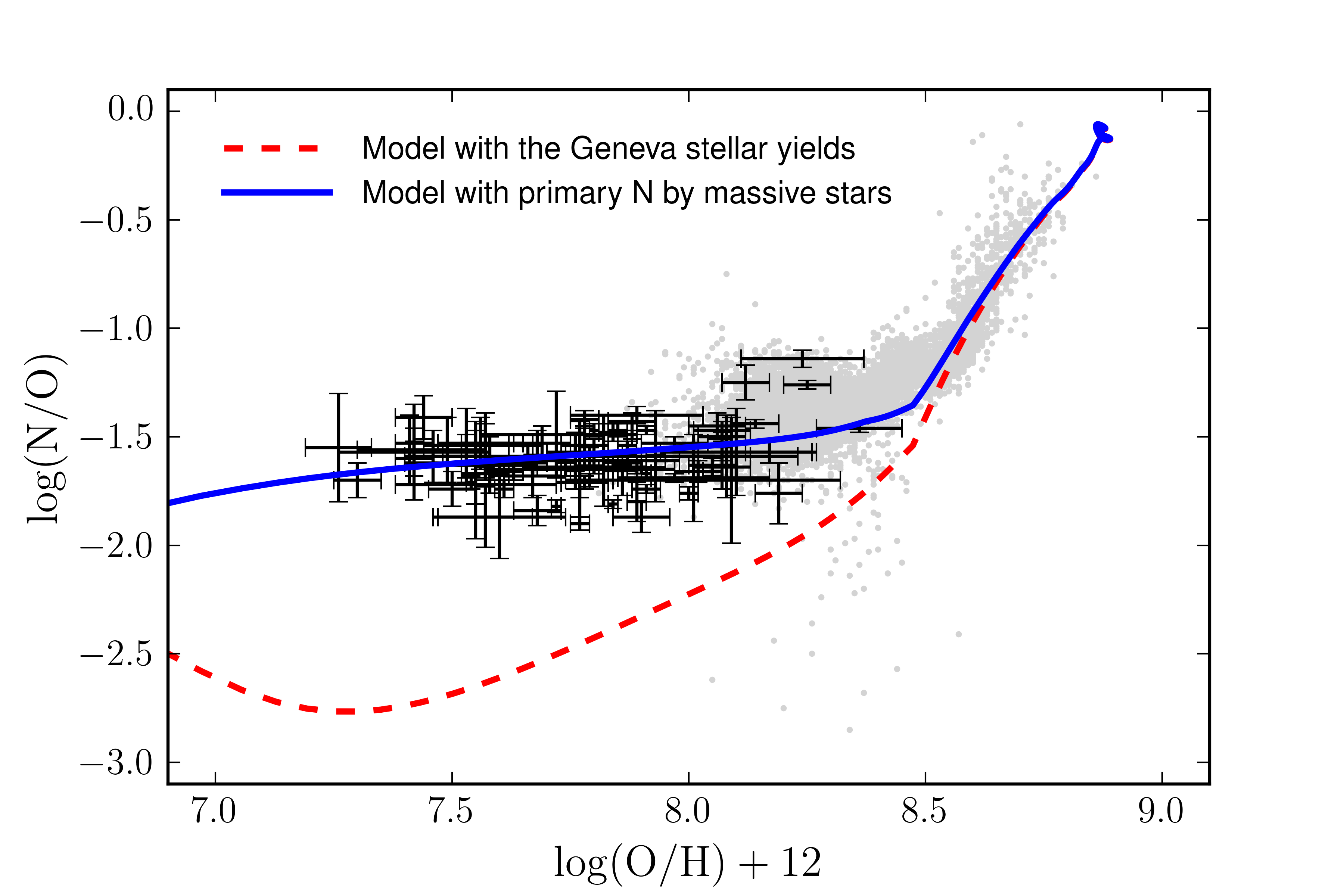}
\caption{ In this figure, we compare the predictions of a model with pure primary N production 
by massive stars with a similar model assuming the Geneva stellar yields for massive stars, as given in \citet[their model 15]{romano2010}. 
 Both models assume $\nu=1.5\,\text{Gyr}^{-1}$, $M_{\mathrm{inf}}=10^{9}\,\mathrm{M}_{\sun}$ and $\tau_\mathrm{inf}=0.1\,\mathrm{Gyr}$. The 
 data are same as in Fig. \ref{fig:dwarf}.
 }
 \label{fig:dwarf_originalN}
\end{figure} 


\item The range in metallicity of the initial (N/O) plateau in the (N/O) vs. (O/H) abundance diagram is determined by the SFE; in particular, 
the higher the SFE, the larger is the extension in metallicity of the plateau. In fact, the SFE is the main parameter driving the rate of metal production from massive stars, 
hence regulating the metallicity of the system when LIMS 
begin to heavily pollute the ISM with nitrogen. Therefore, if the SFE is high, the change in slope of the (N/O) ratio occurs at higher metallicity than in the case with low SFE. 
This is a consequence of the so-called time-delay model on the (N/O) vs. (O/H) diagram. 
In conclusion, the position of the galaxies along the observed (N/O) vs. (O/H) sequence is mostly determined by the SFE. 

\item Only by assuming differential galactic winds, removing exclusively chemical elements produced in core-collapse SNe, we have been able to reproduce the steep increasing trend of 
(N/O) ratios at high metallicity. Nevertheless, the larger amounts of secondary N provided by LIMS as the metallicity increases is necessary to reach the 
high observed (N/O) ratios at high metallicity. In our reference model, we have assumed a mass loading factor of the order of the unity; 
the variation of this parameter crucially determines the slope with which the (N/O) ratios are observed to increase at high metallicity. 
On the other hand, all our models with normal galactic wind -- in which all the chemical elements are carried out of the galaxy potential well 
with the same efficiency -- fail in explaining the observed trend of the (N/O) vs. (O/H) abundance pattern.

\item The role of the IMF consists in giving different weights to stars as functions of their mass, when the star formation process takes place. 
If the IMF is rich in massive stars, then an enhanced O production is predicted (see also \citealt{vincenzo2015}), letting the (N/O) ratios start to increase at high 
(O/H) abundances. So the main effect of the IMF is to shift the (N/O) vs. (O/H) relations over the (O/H) axis. 
Our chemical evolution models with the \citet{kroupa1993} IMF provide the best agreement with the observed dataset.

\end{enumerate}

\section*{Acknowledgements}

FV thanks the Cavendish Astrophysics Group at the University of Cambridge for kindly supporting his visit 
during September 2014. 
FB acknowledges funding from the United Kingdom Science and Technology Facilities Council (STFC). 
RM acknowledges funding from the United Kingdom STFC through grant ST/M001172/1. 
FM acknowledges financial support from PRIN-MIUR~2010-2011 project 
`The Chemical and Dynamical Evolution of the Milky Way and Local Group Galaxies', prot.~2010LY5N2T. 
We thank an anonymous referee for his/her constructive comments, which have improved the clarity of the paper. 
The data used in this paper can be retrieved at the
MPA-JHU DR7 release of spectrum measurements web page:
\url{http://wwwmpa.mpa-garching.mpg.de/SDSS/DR7/}. 


\bsp

\label{lastpage}

\end{document}